\documentclass[conference,compsoc]{IEEEtran}

\ifCLASSOPTIONcompsoc
  \usepackage[nocompress]{cite}
\else
  \usepackage{cite}
\fi

\usepackage{amsfonts,amsmath,amssymb,amstext,latexsym}
\usepackage{balance}
\usepackage{enumerate}
\usepackage{graphicx}
\usepackage{subcaption}
\usepackage{ifpdf}
\usepackage{ifthen}
\usepackage{multirow}
\usepackage[all]{nowidow}
\usepackage{color}
\usepackage[usenames,dvipsnames,table]{xcolor}
\usepackage{xspace}

\newcommand{\thetitle}{Same Same, but Different: A Descriptive Differentiation of Intra-cloud Iaas Services}
\newcommand{\theauthors}{Yehia Elkhatib, Faiza Samreen, Gordon S. Blair}
\newcommand{\thekeywords}{Cloud computing, Cloud brokers}
\ifpdf
	\usepackage[
		colorlinks=true,%
		linkcolor=black,%
		citecolor=black,%
		filecolor=black,%
		urlcolor=black,%
		pdftex,%
		pdftitle={\thetitle},%
		pdfauthor={\theauthors},%
		pdfkeywords={\thekeywords{}, \theauthors},%
		unicode]{hyperref}
\fi

\newcommand*{\cf}{\textit{cf.}\@\xspace}
\newcommand*{\eg}{\textit{e.g.}\@\xspace}
\newcommand*{\ie}{\textit{i.e.}\@\xspace}
\makeatletter
\newcommand*{\etc}{%
    \@ifnextchar{.}%
        {etc}%
        {etc.\@\xspace}%
}
\newcommand*{\etal}{%
    \@ifnextchar{.}%
        {et al}%
        {et al.\@\xspace}%
}
\makeatother

\newboolean{showcomments}
\setboolean{showcomments}{false} 
\ifthenelse{\boolean{showcomments}}
{ \newcommand{\mynote}[3]{
    \fbox{\bfseries\sffamily\scriptsize#1}
    {\small$\blacktriangleright$\textsf{\emph{\color{#3}{#2}}}$\blacktriangleleft$}}}
{ \newcommand{\mynote}[3]{}}
\newcommand{\shrink}[1]{}

\newcommand{\fig}[1]{Figure~\ref{#1}}
\graphicspath{{./figs/}}
\newcommand{\figwidth}{0.99\columnwidth}

\begin{document}
\title{\thetitle}

\author{
	\IEEEauthorblockN{\theauthors}\\
	\IEEEauthorblockA{MetaLab, School of Computing and Communications, Lancaster University, UK\\
	Email: \{i.lastname\}@lancaster.ac.uk}\\
}

\maketitle

\begin{abstract}
Users of cloud computing are overwhelmed with choice, even within the services offered by one provider. As such, many users select cloud services based on description alone. We investigate the degree to which such strategy is optimal. In this quantitative study, we investigate the services of 2 of major IaaS providers. We use 2 representative applications to obtain longitudinal observations over 7 days of the week and over different times of the day, totalling over 14,000 executions. We give evidence of significant variations of performance offered within IaaS services, calling for brokers to use automated and adaptive decision making processes with means for incorporating expressive user constraints. 
\end{abstract}

%
\IEEEpeerreviewmaketitle

\section{Introduction}

The cloud is a transformative computing paradigm that has touched almost every application in the modern world. The cloud computing market is a fierce one with high competition between enormous multinational technology companies such as Google, Amazon and IBM, as well as 
more specialised companies such as Flexiant and DigitalOcean. There are well documented differences between such cloud service providers (CSPs), most notably in terms of pricing schemes and hardware heterogeneity. This gives impetus for the development of inter-CSP brokers, an area of active research (\cf~\cite{Javed201652,Quarati2016403,Aazam2017}). 

However, there is also need for work on \emph{intra}-CSP decision support. On the surface of it, the services offered by any single CSP might seem straight forward as they are classified under easily identifiable tags such as \emph{general-purpose}, \emph{high-memory}, and \emph{cpu-optimised}. As such, intuition would dictate that a customer simply needs to select a class that matches their application type and then select an instance from that class that falls within their budget.

\begin{figure}[h]
    \centering
    \includegraphics[width=0.9\columnwidth]{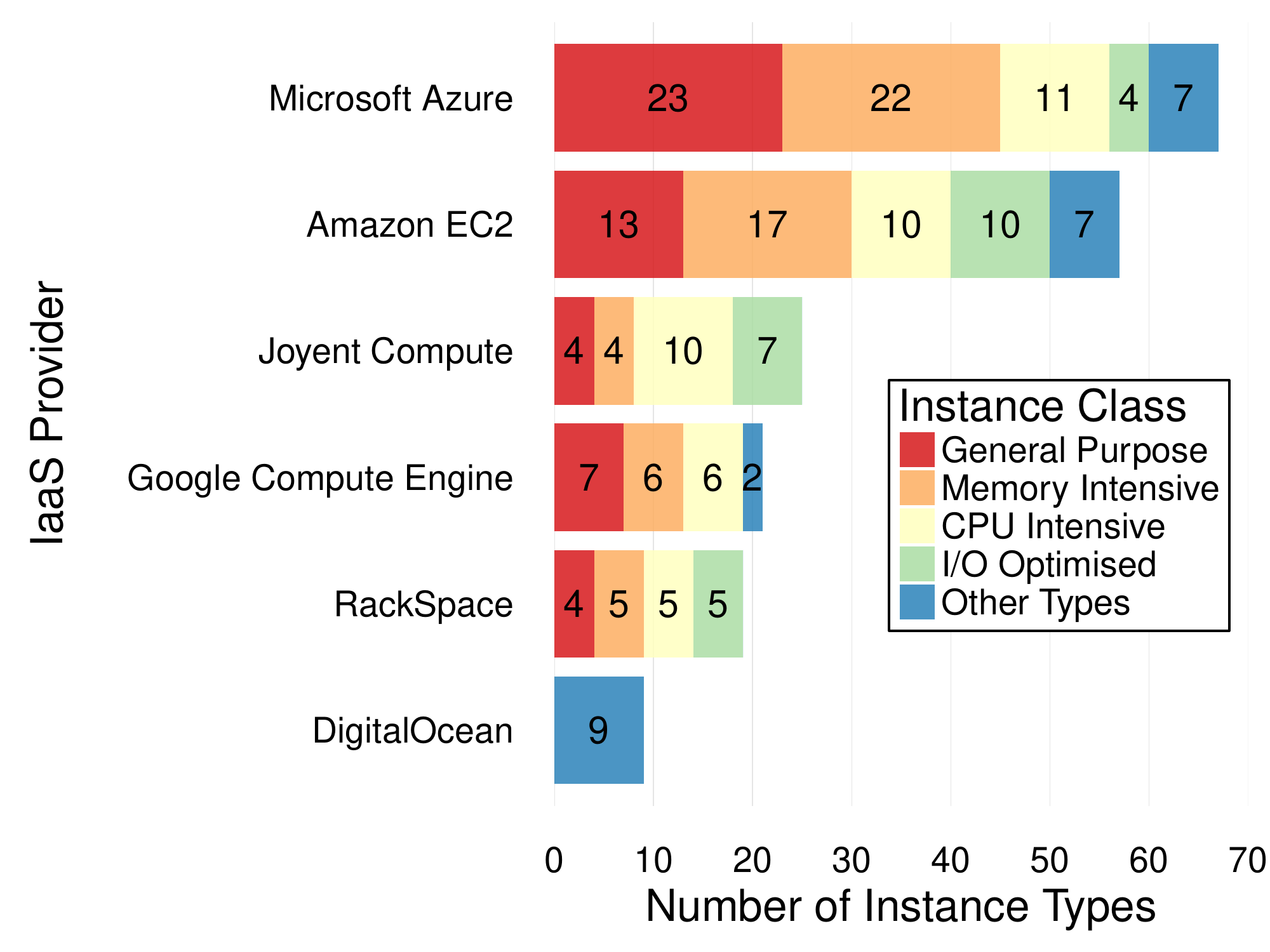}
	\caption{The number of Linux-based instance types offered by major IaaS vendors, as of July 2017.}
	\label{fig:instance-types}
\end{figure}

Nonetheless, selection is not as easy as it looks. First, most CSPs offer a bewildering range of IaaS services in the form of different instance types under some variants of the aforementioned classes. A quick survey of the major CSPs demonstrates this as depicted in \fig{fig:instance-types}. CSPs such as Amazon and Microsoft offer a total of 57 and 67 instance types, respectively. One CSP not represented in the plot is IBM SoftLayer, which allows its customers to create custom instances using parameter sliders including number of cores (between 1 and 56), memory (between 1GB and 242GB), and storage (between 25GB and 100GB), as well as other settings. In total, this offers a customer to select from a range of 768 possible permutations!

Second, and more importantly for our purposes in this paper, earlier work \cite{Samreen2016Daleel,leitner2016patterns} has indicated that instances of a single CSP are not necessarily equally cost effective. In other words, IaaS customers do not get more performance the more money they pay, even within instances of the same class (\eg high-memory) of a single CSP.

In this paper, we investigate variation between intra-CSP instance types in detail. For this, we employ intensive benchmarking using 2 representative real-world applications, running them 14,300+ times over different times and days of the week. We use this methodology to analyze the intra-CSP performance of 2 major IaaS CSPs, namely Amazon Elastic Compute Cloud (EC2) and Google Compute Engine (GCE).

Our philosophy here is pragmatic. We use real applications that are representative and are open source, thus allowing others to produce follow-up comparative studies (as rightly called for in \cite{leitner2016patterns}). We also take the user's perspective, observing application performance as a user would. 

The rest of the paper is organised as follows. We discuss related work (\S\ref{sec:rw}). Next, we explain our measurement methodology and describe the applications we employ as use cases (\S\ref{sec:meth}). Then we present and describe the results (\S\ref{sec:results}) and comment on their wider implications (\S\ref{sec:disc}). 

\section{Related Work}
\label{sec:rw}

It is well known that cloud services are plentiful and varied. As a result, cloud computing literature is crowded with efforts attempting to optimise service selection using a number of different approaches. 
However, the vast majority of this work is flawed as they rely on the ``book value'' attributed to the services on offer which is not necessarily representative or reliable. An early study \cite{Ostermann2010} looked into variances between a few EC2 instance types using standard benchmarks and noted that there are no ``best performing instance type''. 
Other studies have identified variances within one provider \cite{Ostermann2010,5708447,Li2013,7027861}, and over a span of several days \cite{6655674,leitner2016patterns}, different times of the year \cite{Schad2010}, and different regions of a single IaaS provider \cite{lucas2013multi,leitner2016patterns}. 
Recent works \cite{Samreen2016Daleel,leitner2016patterns} indicated that cloud instance performance is difficult to foresee based on the information the CSP offers and, thus, selecting the optimal instance is a non-trivial decision.

Therefore, some have tried to gain better understanding of the potential performance of cloud instances using standard or bespoke benchmarking suites and various modelling techniques; \eg \cite{6133223,6753834,7027861,7037674}. Others use profile-based methods (\cf \cite{Cloudcmp,Verghese2013Cloudbench,cherrypick2017}) or application-specific performance models (\cf \cite{Venkataraman2016Ernest}). 

However, little attention was given to variation in cost-effectiveness over similar instances or of the performance of a single instance type. Consequently, our knowledge of the IaaS instance search space is still constrained in dimensionality. This is what we address in our study.

\section{Methodology}
\label{sec:meth}
We now detail our experimental objectives, strategy, use cases, and measurement approach.

\subsection{Objectives}
The objectives of our experimental strategy are to:
\begin{itemize}
    \item Ascertain if instance classification and description is indeed helpful for instance selection or not.
    \item Identify the degree of variation in the performance of a single instance type.
    \item Uncover differences in cost effectiveness between instance types of a single CSP.
\end{itemize}

\subsection{Experimental Strategy}
Our overall strategy is to study the variation in the performance of running a uniform application workload over different instance types. 
In order to collect enough data points to identify any potential performance variance, we repeated the workload with a 10 minute delay between each pair of runs. 

All application parameters and input were kept constant between application runs in order to reduce the dimensionality of the experiments. 

For example, VARD requires external file as an input and the experiments were conducted using typical input files of fixed size of 3kb. While running the smallpt application, the grid size to render the image was set to 200.
We also ran the applications over different times of the day and over all days of one week to control for temporal variances such as diurnal patterns. 

\subsection{Cloud Infrastructures}
We identified our target infrastructures as Amazon EC2 and Google GCE as two of the major players in the IaaS market. 
From each, we examined a subset of their instance types that seem suitable for running each of our applications. Note that there hardly was a straightforward answer to ``which instance type is best for running this application'', which is the point of running this study. As such, we ended up with a set from each provider. These are summarised in Tables~\ref{tab:instance-types-ec2}--\ref{tab:instance-types-gce}, accurate as of the time of the experiments. All instances were running 64-bit Ubuntu Linux version 14.04.

\begin{table}[!ht]
\def\arraystretch{1.2}
\setlength{\tabcolsep}{3pt}
\centering
\caption{Computational specifications of EC2 instances.}
\begin{tabular}{ccccccc}
	\hline
	\textbf{Series} & \textbf{Instance} & \textbf{vCPU} & \textbf{ECU} & \textbf{RAM} & \textbf{Storage} & \textbf{Price}\\
	&\textbf{Type}&&&(GiB)&(GB)& (\$/h)\\
	\hline
	T2 (General & \texttt{t2.small} & 1 & Var. & 2 & 20 & 0.026\\
	\cline{2-7}
	Purpose) & \texttt{t2.medium} & 2 & Var. & 4 & 20 & 0.052\\
	\hline
	M3 & \texttt{m3.medium} & 1 & 3 & 3.75 & 4(S) & 0.070\\
	\cline{2-7}
	(General & \texttt{m3.large} & 2 & 6.5 & 7.5 & 32(S) & 0.140\\
	\cline{2-7}
	Purpose) & \texttt{m3.xlarge} & 4 & 13 & 15 & 32(S) & 0.280\\
	\hline
	C4 & \texttt{c4.large} & 2 & 8 & 3.75 & 20 & 0.116\\
	\cline{2-7}
	(Compute & \texttt{c4.xlarge} & 4 & 16 & 7.5 & 20 & 0.232\\
	\cline{2-7}
	Optimised) & \texttt{c3.xlarge} & 4 & 14 & 7.5 & 32(S) & 0.239\\
	\hline
\end{tabular}
\label{tab:instance-types-ec2}
\end{table}

\begin{table}[!ht]
\def\arraystretch{1.2}
\setlength{\tabcolsep}{3pt}
\centering
\caption{Computational specifications of GCE instances.}
\begin{tabular}{ccccccc}
	\hline
	\textbf{Series} & \textbf{Instance} & \textbf{vCPU} & \textbf{GCEU} & \textbf{RAM} & \textbf{Storage} & \textbf{Price}\\
	&\textbf{Type}&&&(GB)&(GB)& (\$/h)\\
	\hline
	Standard & \texttt{n1-standard-1} & 1 & 2.75 & 3.75 & 16 & 0.036\\
	\cline{2-7}
	Type & \texttt{n1-standard-2} & 2 & 5.5 & 7.5 & 16 & 0.071\\
	\cline{2-7}
	& \texttt{n1-standard-4} & 4 & 11 & 15 & 16 & 0.142\\
	\hline
	High Mem. & \texttt{n1-highmem-2} & 2 & 5.5 & 13 & 16 & 0.106\\
	\hline
	High & \texttt{n1-highcpu-2} & 2 & 5.5 & 1.8 & 16 & 0.056\\
	\cline{2-7}
	CPU & \texttt{n1-highcpu-4} & 4 & 11 & 3.6 & 16 & 0.118\\
	\cline{2-7}
	& \texttt{n1-highcpu-8} & 8 & 2.2 & 7.2 & 16 & 0.215\\
	\hline
\end{tabular}
\label{tab:instance-types-gce}
\end{table}

Only on-demand instances are used for our experiments. These have no long-term commitments and are charged on a pay-as-you-go basis at an hourly rate. All instances are chosen to be located in western Europe zones, which are hosted in data centers in Ireland. 
The `Price' column refers to the hourly charge for running a VM of the referenced instance type.

It is important to note the units the two providers use to describe their respective instance types. Both give the number of virtual cores assigned to a VM (\ie `vCPU'). They both give an indicative amount of CPU capacity, but they each use their own opaque unit: EC2 uses `ECU' while GCE uses `GCEU'. 
Amazon does not advise how an EC2 Compute Unit (ECU) relates to physical processing speed; it only assures that it is a standard unit across its different IaaS offerings. 
Google compute engine unit (GCEU)\footnote{Pronounced as GQ.} is an abstraction of compute resources where, according to Google, 2.75 GCEUs represent the minimum power of one logical core. 
In either case, there is no clear indication how they relate to physical processing speed. 


\subsection{Use Cases}
\label{sec:apps}

We have selected 2 representative applications with different  architectures and categories relating to their intensity of memory and CPU usage.

\subsubsection{VARD}
VARD~\cite{baron2008vard2} is a tool designed to detect and tag spelling variations in historical text, particularly those written in Early Modern English. The output is aimed at improving the accuracy of other corpus analysis solutions. Hence, VARD is considered a pre-processor tool to a wide range of corpus linguistic tools such as NLP analysis, semantic tagging, annotations, \etc. An example use is to detect misspelled words in SMS~\cite{Tagg2009}.

VARD runs as a single threaded Java application. It is memory intensive as it holds in memory a representation of the full text, as well as various dictionaries that are used for normalising spelling variations and which are constantly being updated as text is being processed. 
VARD is representative of applications used for various other uses such as business transaction processing, document analysis, and web app (\eg mobile gaming) backends.

\subsubsection{smallpt} 
smallpt is a popular open source C++ application for rendering indirect illumination in 3D graphical scenes. In effect, it simulates multiple light sources and how their illumination reflects off different objects in a three dimensional space. smallpt does this using a method called unbiased Monte Carlo path tracing, which is an expensive simulation of light paths.

smallpt is a multi-threaded OpenMP based application, and is categorised as CPU intensive. OpenMP is used to achieve parallelism for dynamic allocation of rows of the image to different threads where each thread is allocated to an available core.
The smallpt application is a composition of different features such as anti aliasing, ray-sphere  intersection and Russian roulette for path termination. It requires a number of samples per pixel as input, which is considered as number of paths per pixel for rendering a scene.

For our benchmarking, we selected a box scene that is constructed out of nine very large overlapping spheres. The image is computed using equations that solve the rendering equation. The Monte Carlo path tracing algorithm is used with Russian roulette for path termination.



\subsection{Measurement Setup}
Our main performance metric is the time an application workload takes to execute. 
In addition, we also use a suite of Linux system performance monitoring tools (namely \texttt{vmstat}, \texttt{glances}, and \texttt{sysstat}) for continuously monitoring VM resource utilisation.

\section{Results}
\label{sec:results}

We now describe the outcome of our experiments. We start with the overall distributions, and then move on to explore the cost-effectiveness of different instance types.

\subsection{Overall Distributions}
\label{sec:results:overall}

\begin{figure*}[tbh]
    \centering
    \begin{subfigure}{\figwidth}
        \includegraphics[width=\figwidth, trim=0cm 0.2cm 0cm 0.2cm, clip]{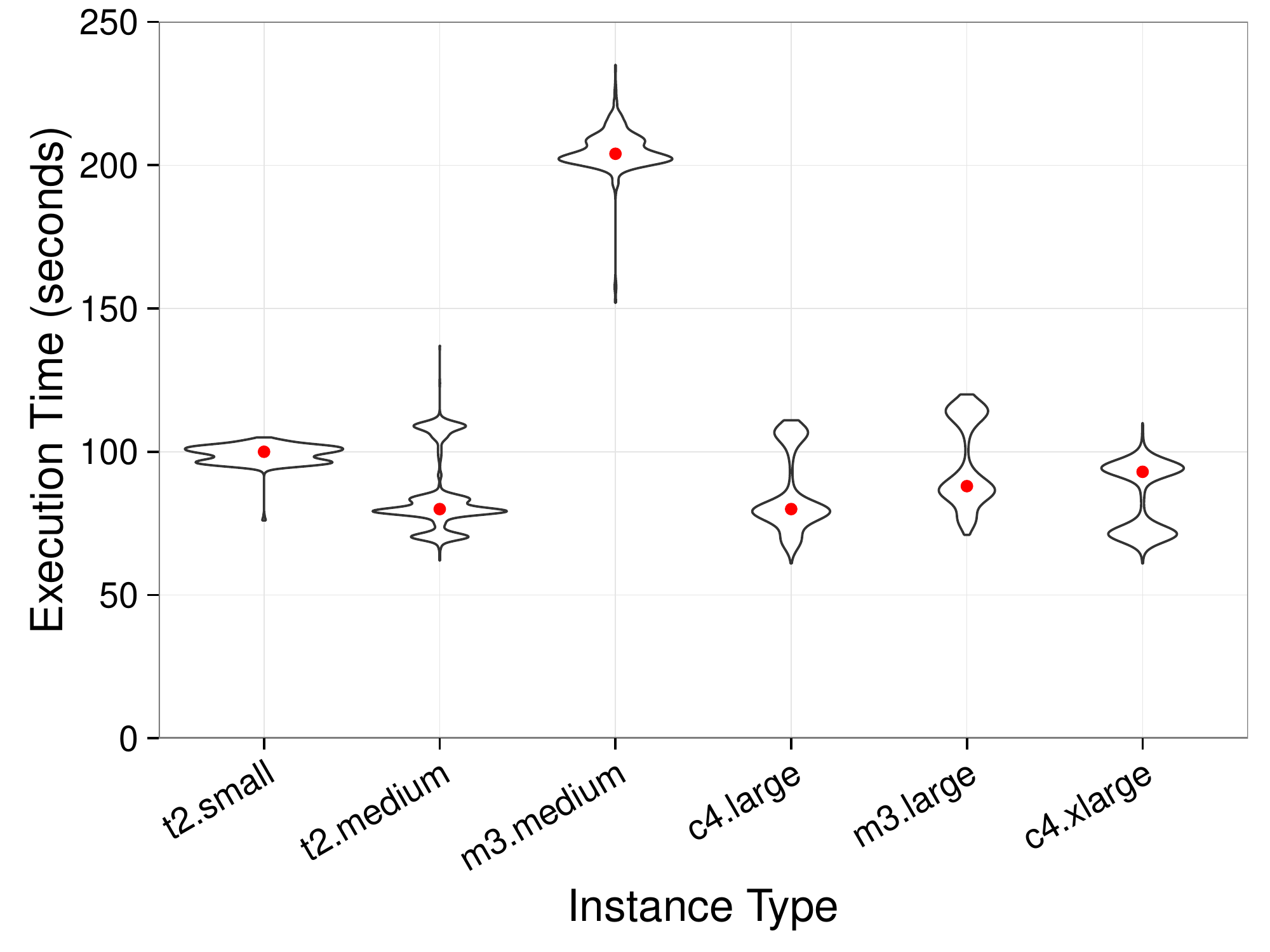}
        \caption{EC2}
	\end{subfigure}\hfill
    \begin{subfigure}{\figwidth}
        \includegraphics[width=\figwidth, trim=0cm 0.2cm 0cm 0.2cm, clip]{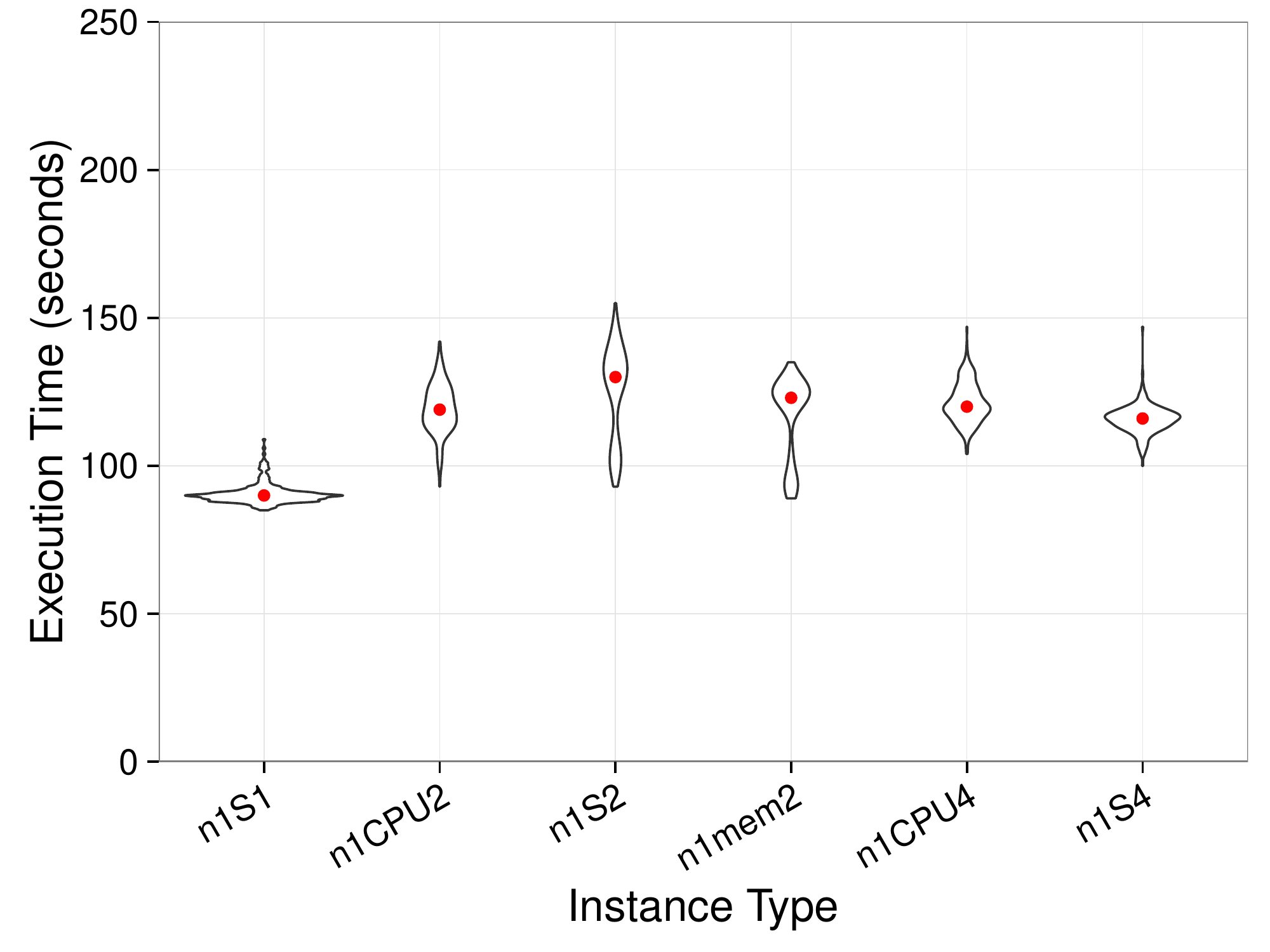}
        \caption{GCE}
	\end{subfigure}\par\medskip
	\caption{The overall distribution of VARD execution times.\label{fig:VARD-violins}}
\end{figure*}

\begin{figure*}[tbh]
    \centering
    \begin{subfigure}{\figwidth}
        \includegraphics[width=\figwidth, trim=0cm 0.2cm 0cm 0.2cm, clip]{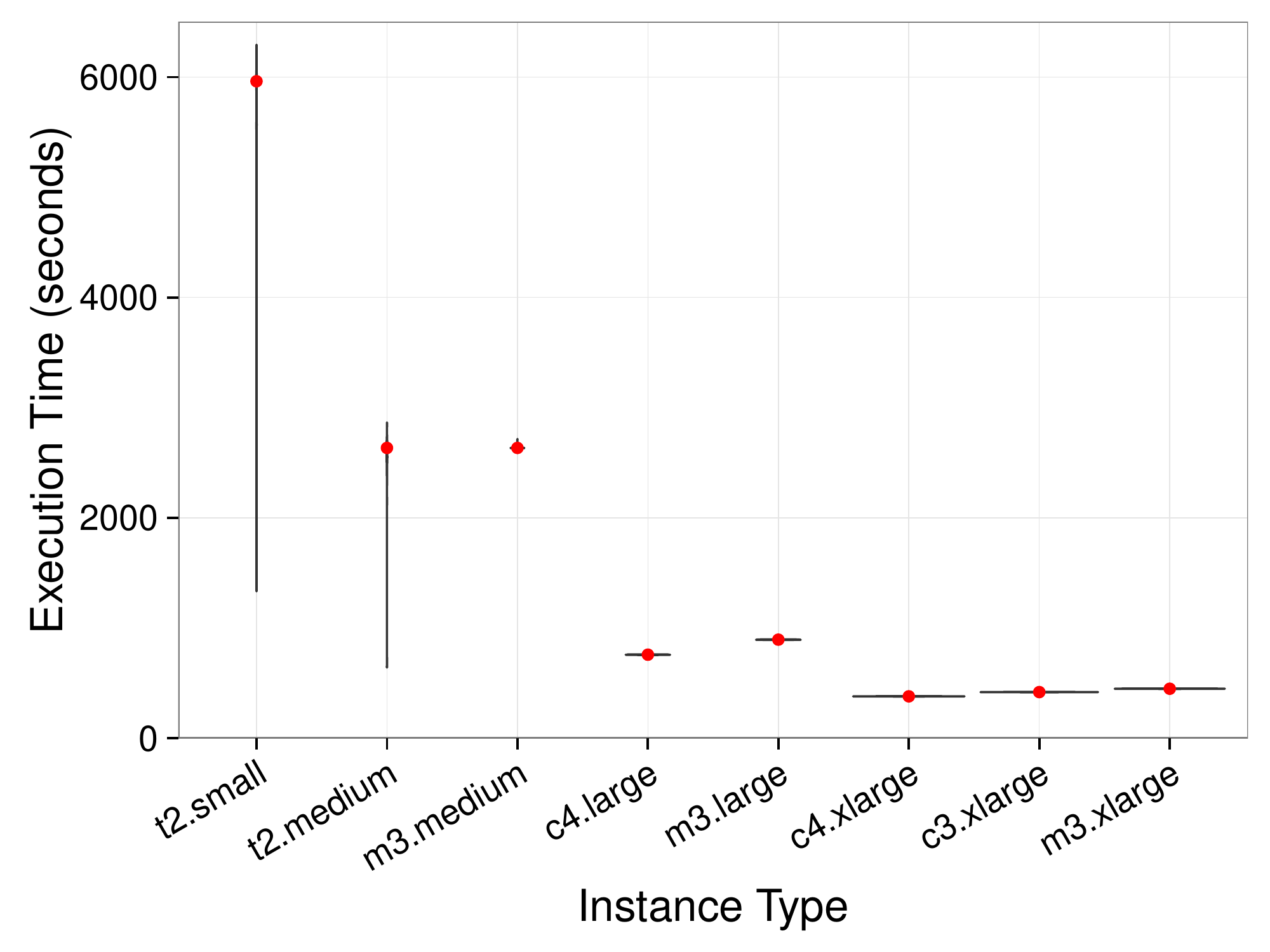}
        \caption{EC2}
	\end{subfigure}\hfill
    \begin{subfigure}{\figwidth}
        \includegraphics[width=\figwidth, trim=0cm 0.2cm 0cm 0.2cm, clip]{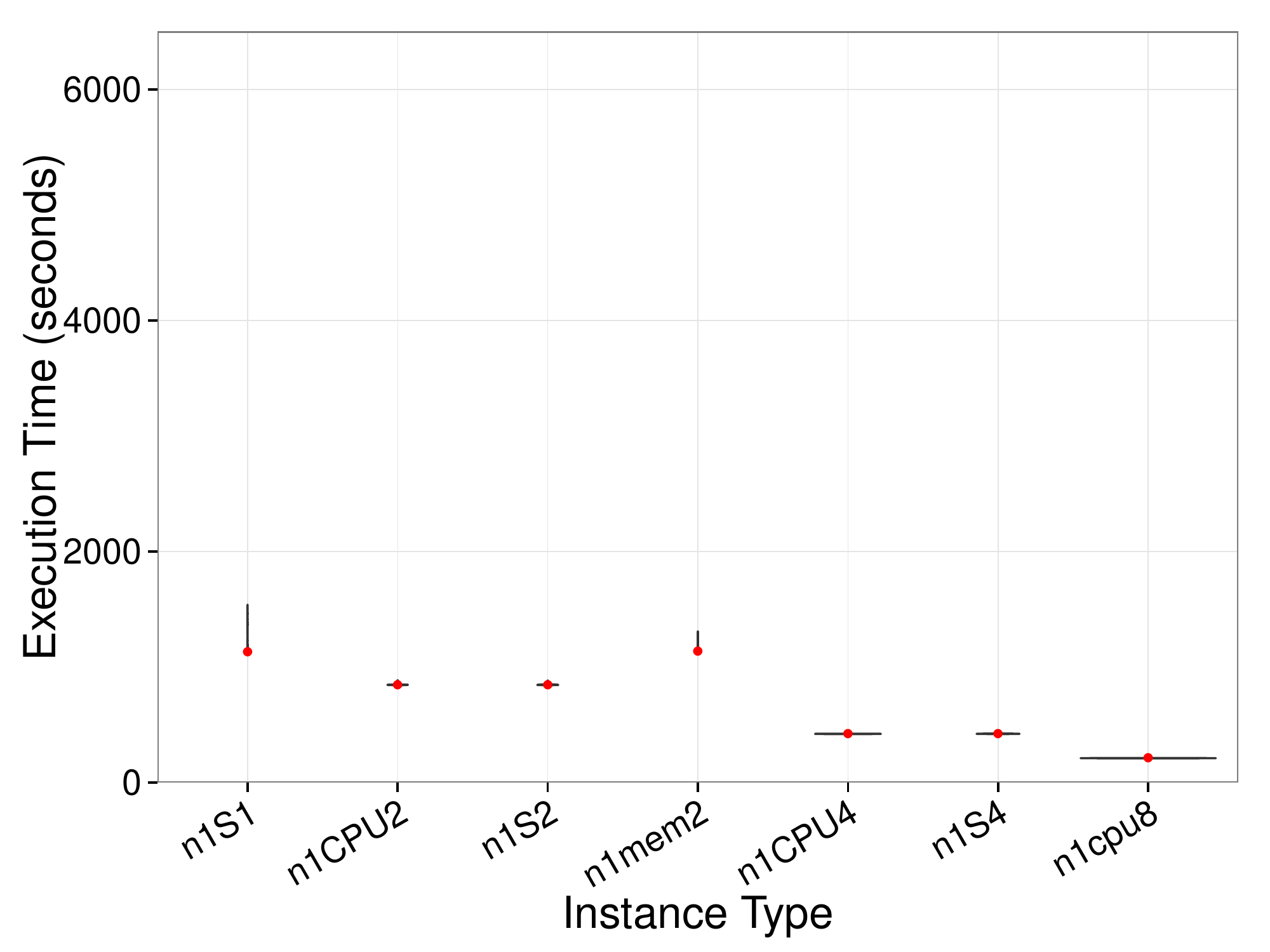}
        \caption{GCE}
	\end{subfigure}\par\medskip
	\caption{The overall distribution of smallpt execution times.\label{fig:smallpt-violins}}
\end{figure*}

The distributions of application execution times are displayed using violin plots, where red dots mark the median: \fig{fig:VARD-violins} for VARD, and \fig{fig:smallpt-violins} for smallpt. Note that all instances types are sorted increasingly by cost from the left hand side. 
From these plots, we immediately observe some interesting and contrasting patterns.

First, running smallpt is generally much more predictable in terms of how long it will take to execute successfully than VARD. Quartiles are narrow for smallpt over most instances types. We ascribe this to smallpt being a CPU intensive application, as memory is typically under higher contention from other guest VMs than CPU resources are. 
This is validated when we inspect the only exceptions to the observation being discussed: the cheapest 2 instance types on EC2, which exhibit large uncertainty. This is easily explained considering EC2's CPU Credits scheme\footnote{\url{http://docs.aws.amazon.com/AWSEC2/latest/UserGuide/t2-instances.html\#t2-instances-cpu-credits}}, which is offered only on the \texttt{T2} instance series. Under this scheme, customers collect credits for idle instances that they can later spend for full CPU utilisation. 

Second, VARD execution times exhibit rather wide ranges. Furthermore, many VARD execution times follow bimodal distributions. This is quite pronounced in some cases, especially in the case of VARD running over EC2 instances. In comparison, GCE instances are again more predictable with distributions that are closer to being uni- than bimodal.

Finally, and perhaps most peculiarly, instance types clearly break the common intuition of ``you get what you pay for''. There are multiple examples of this. One surprising case is that of EC2 \texttt{m3.medium} that consistently performs badly for VARD, a memory intensive application, as well as smallpt. Another example is GCE \texttt{n1S1} outperforming all other GCE instance types for running VARD. 

A general observation from the above confirms the complexity of the task of selecting an IaaS instance. Choosing one purely based on its computational specifications, the performance it promises, or how much budget is available is an assured recipe for uncertainty that may result in extremely low performance, as is the case with EC2 \texttt{m3.medium}. It is worth stressing that this uncertainty is experienced even with EC2 and GCE, 2 of the largest market-leading IaaS providers.

\subsection{What could be done in X hours?}
\label{sec:results:hours}

Due to the observed variation as discussed above, we identify two scenarios that we will use going forward. These are labelled \emph{Best Case} and \emph{Worst Case}, which correspond to the lower and upper quartiles, respectively, as observed in the overall distributions. 

\begin{figure*}[tbh]
    \centering
    \begin{subfigure}{\figwidth}
        \includegraphics[width=\figwidth, trim=0cm 0.2cm 0cm 0cm, clip]{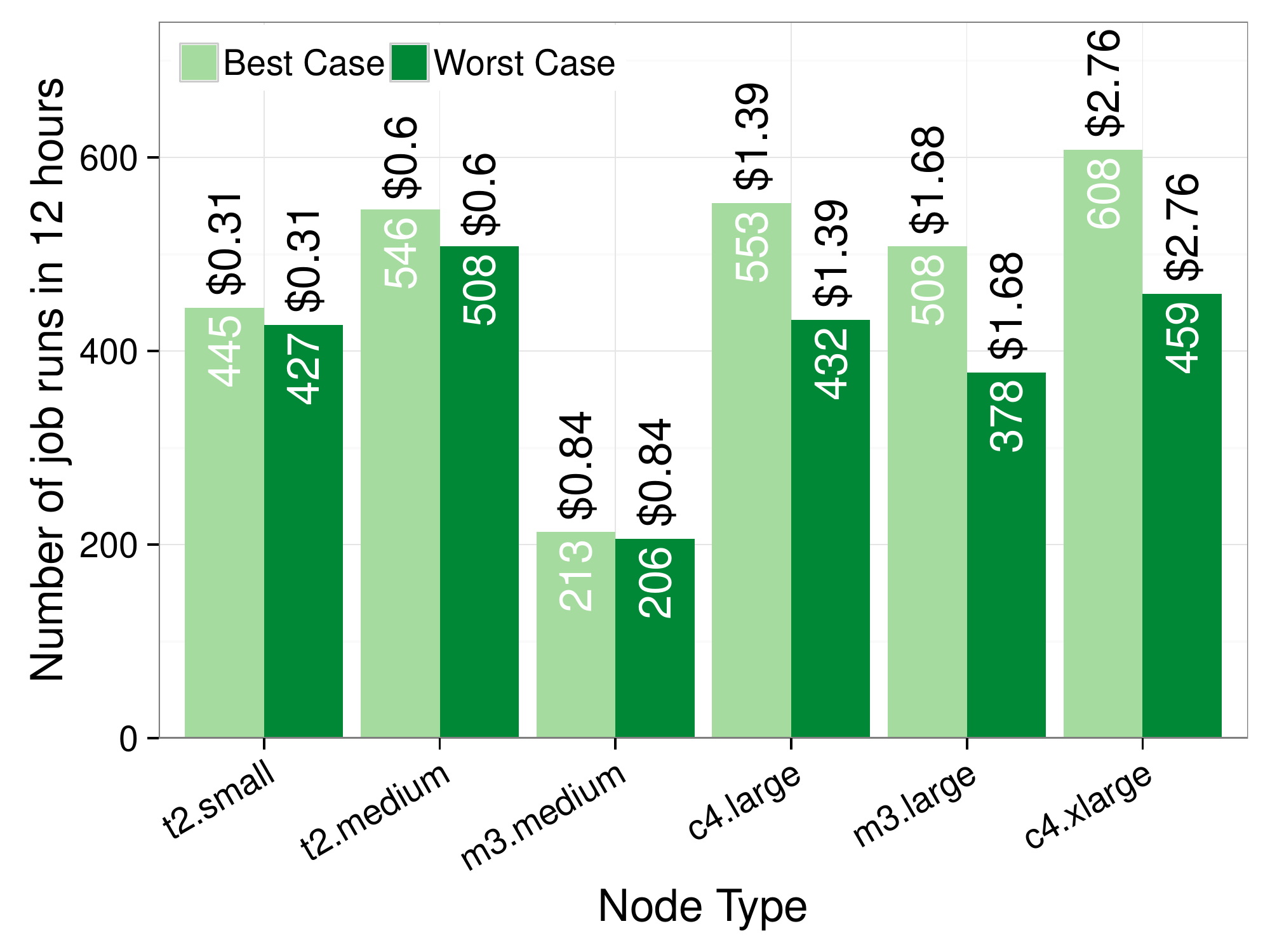}
        \caption{EC2}
	\end{subfigure}\hfill
    \begin{subfigure}{\figwidth}
        \includegraphics[width=\figwidth, trim=0cm 0.2cm 0cm 0cm, clip]{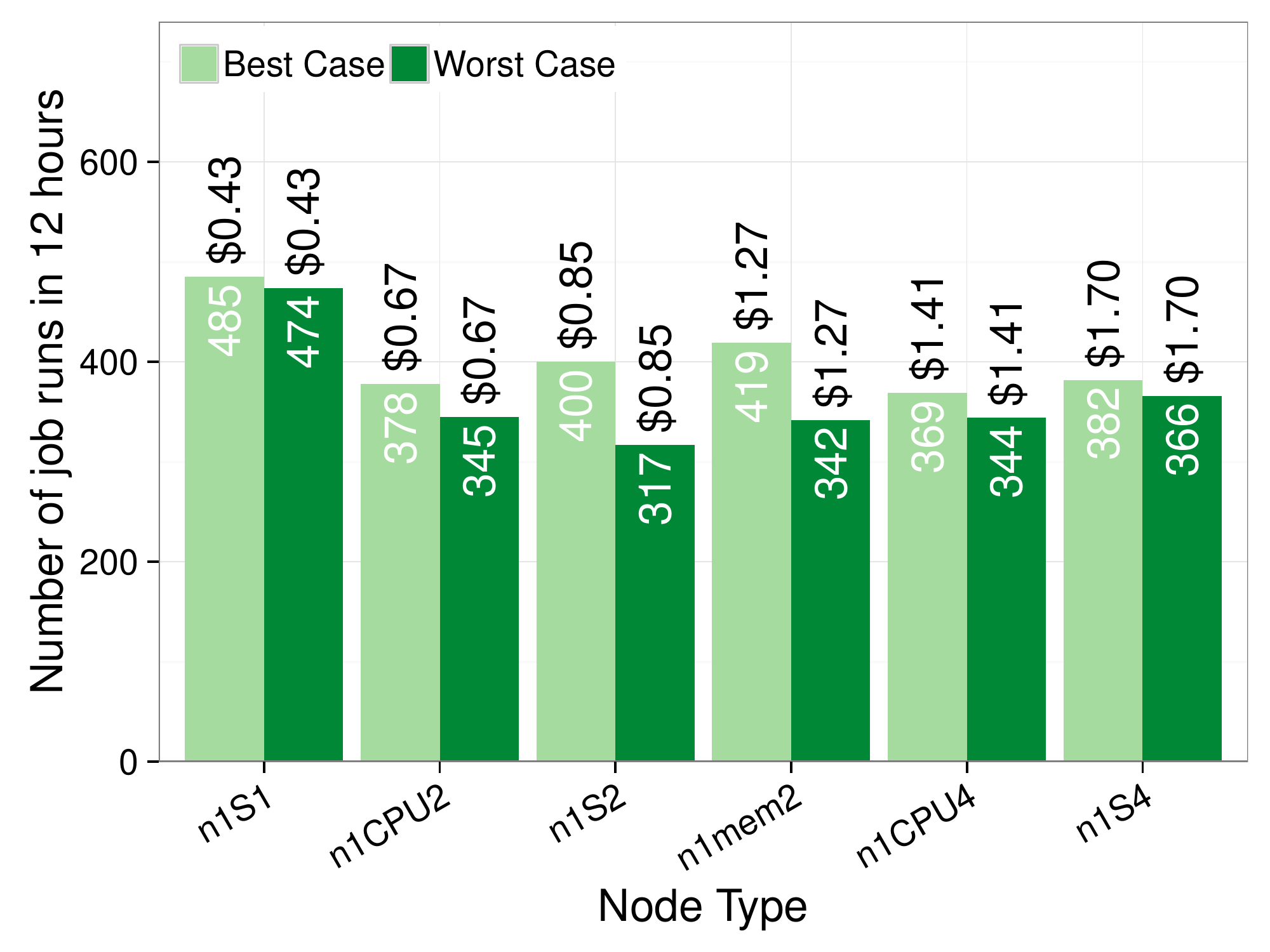}
        \caption{GCE}
	\end{subfigure}\par\medskip
	\caption{The maximum number of VARD job runs in a 12 hour period.\label{fig:vard-12hr}}
\end{figure*}

\begin{figure*}[tbh]
    \centering
    \begin{subfigure}{\figwidth}
        \includegraphics[width=\figwidth, trim=0cm 0.2cm 0cm 0cm, clip]{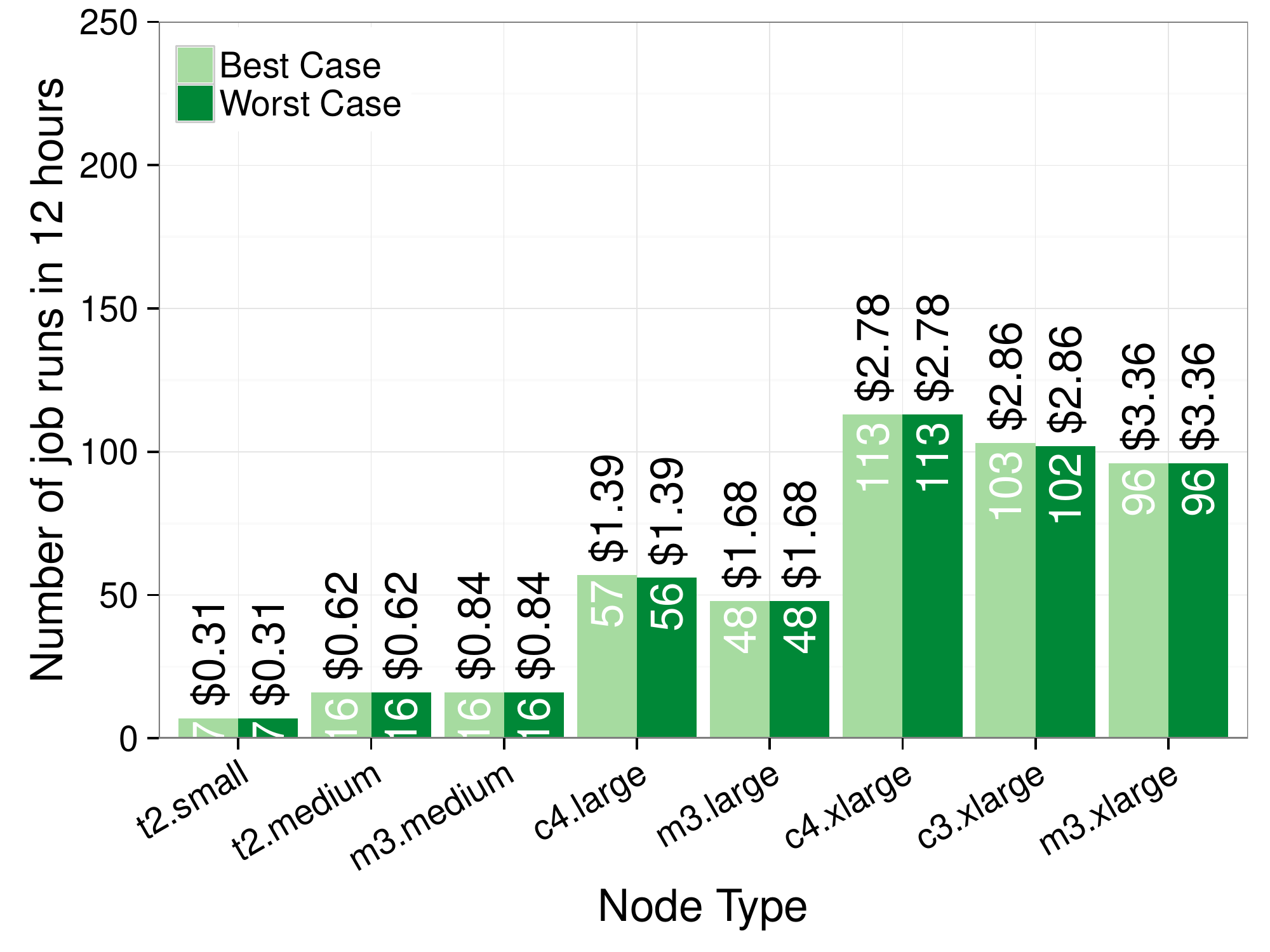}
        \caption{EC2}
	\end{subfigure}\hfill
    \begin{subfigure}{\figwidth}
        \includegraphics[width=\figwidth, trim=0cm 0.2cm 0cm 0cm, clip]{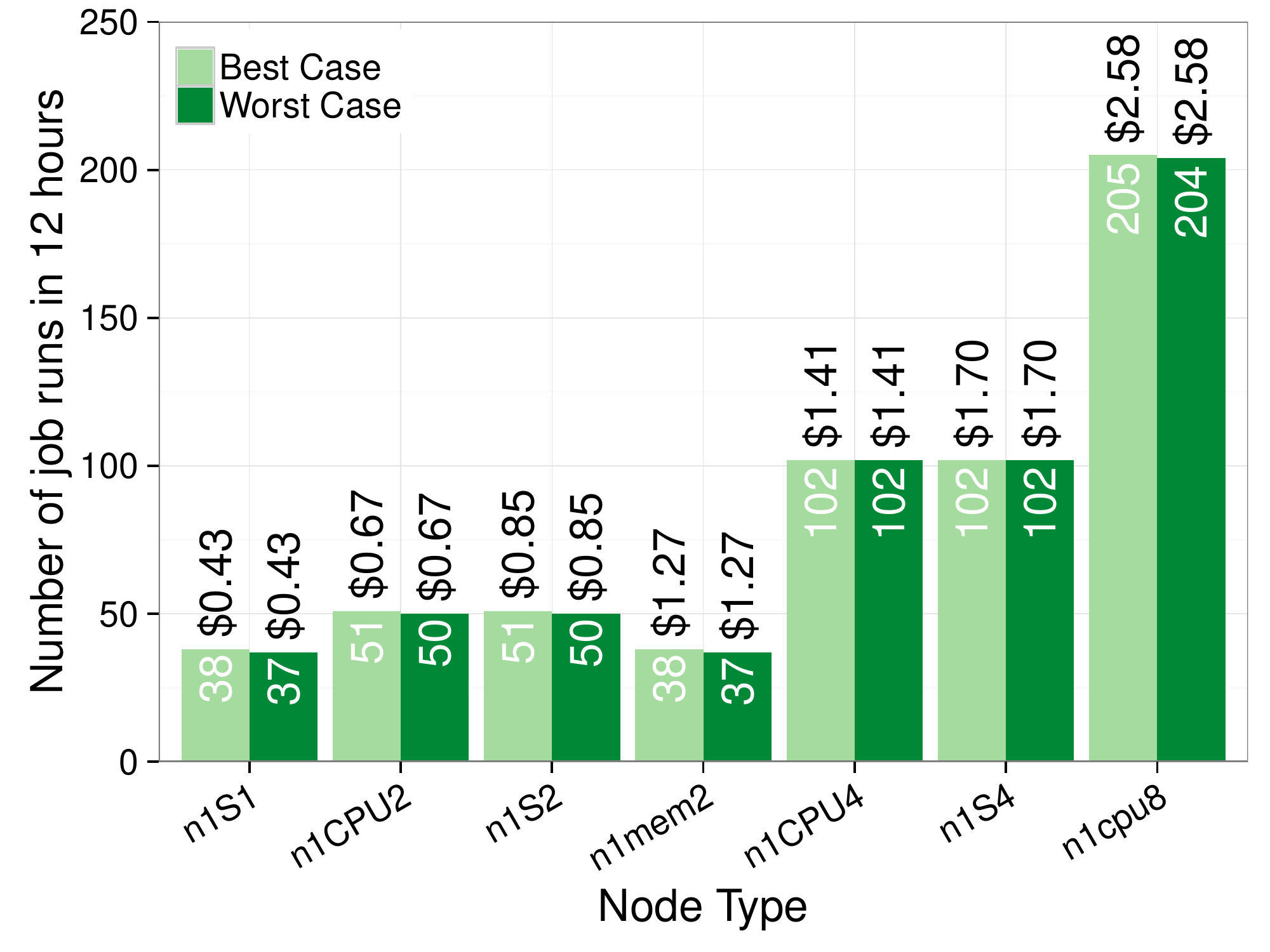}
        \caption{GCE}
	\end{subfigure}\par\medskip
	\caption{The maximum number of smallpt job runs in a 12 hour period.\label{fig:smallpt-12hr}}
\end{figure*}

We focus first on the number of times each application could be executed in a certain amount of time (we chose 12 hours) as a proxy for performance for applications requiring repetitive or Monte Carlo style execution. The corresponding plots are in \fig{fig:vard-12hr} for VARD, and \fig{fig:smallpt-12hr} for smallpt.

The first impression of the plotted results is a reinforcement of the notion that ``you get what you pay for'' does not always hold. This is signified by plots that do not follow the trend of increasing number of executable jobs for more expensive instances. This is especially visible for VARD on both EC2 and GCE. On EC2, one could run just as many jobs on the cheapest instance type (\ie \texttt{t2.small}) as on the most expensive one (\ie \texttt{c4.xlarge}). For GCE, it is actually more effective to use the cheapest instance type (\ie n1S1) than any other. 
With smallpt, the trend is closer to what is expected with some minor deviations. For instance, EC2 \texttt{c4.xlarge} is able to run more jobs than the slightly more expensive \texttt{c3.xlarge} and \texttt{m3.xlarge} instances. 

To examine such counter-intuitive cost-effectiveness further, we calculated the cost of running the maximum number of jobs per instance type. This is indicated by the cost in US\$ above each plotted bar. These costs reveal further implications to instance selection: for relatively short, repetitive workloads like VARD and smallpt, smaller instances are extremely more cost effective than those equipped with higher computational specifications. For example, one could run between 427 and 445 VARD jobs on EC2 \texttt{t2.small} for just \$0.31 as opposed to 459--608 jobs for almost 9 times as much (\$2.76) on \texttt{c4.xlarge}.

Another observation is that VARD performance on higher spec instance types is less certain than on cheaper ones. This was displayed by the general distributions in \S\ref{sec:results:overall}, but is made more evident now with only the interquartile range affecting the best and worst case scenarios. Coupled with the low cost-effectiveness of such instance types as highlighted above, higher spec instance types pose higher risk and cost with relatively less reward.

\subsection{How much (and how long) for Y jobs?}
\label{sec:results:budget}

\begin{figure*}[tbh]
    \centering
    \begin{subfigure}{\figwidth}
        \includegraphics[width=\figwidth, trim=0cm 0.2cm 0cm 0cm, clip]{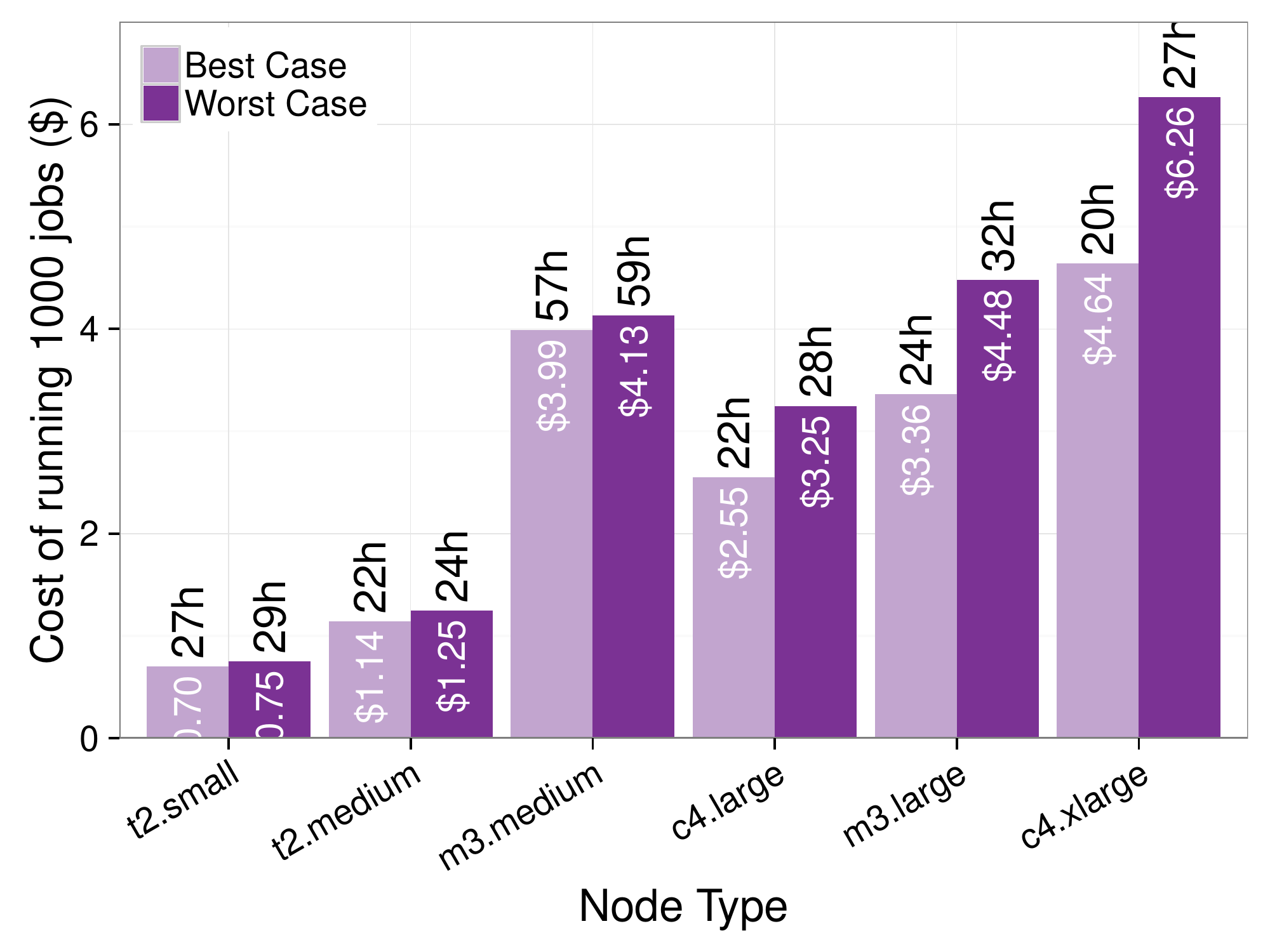}
        \caption{EC2}
	\end{subfigure}\hfill
    \begin{subfigure}{\figwidth}
        \includegraphics[width=\figwidth, trim=0cm 0.2cm 0cm 0cm, clip]{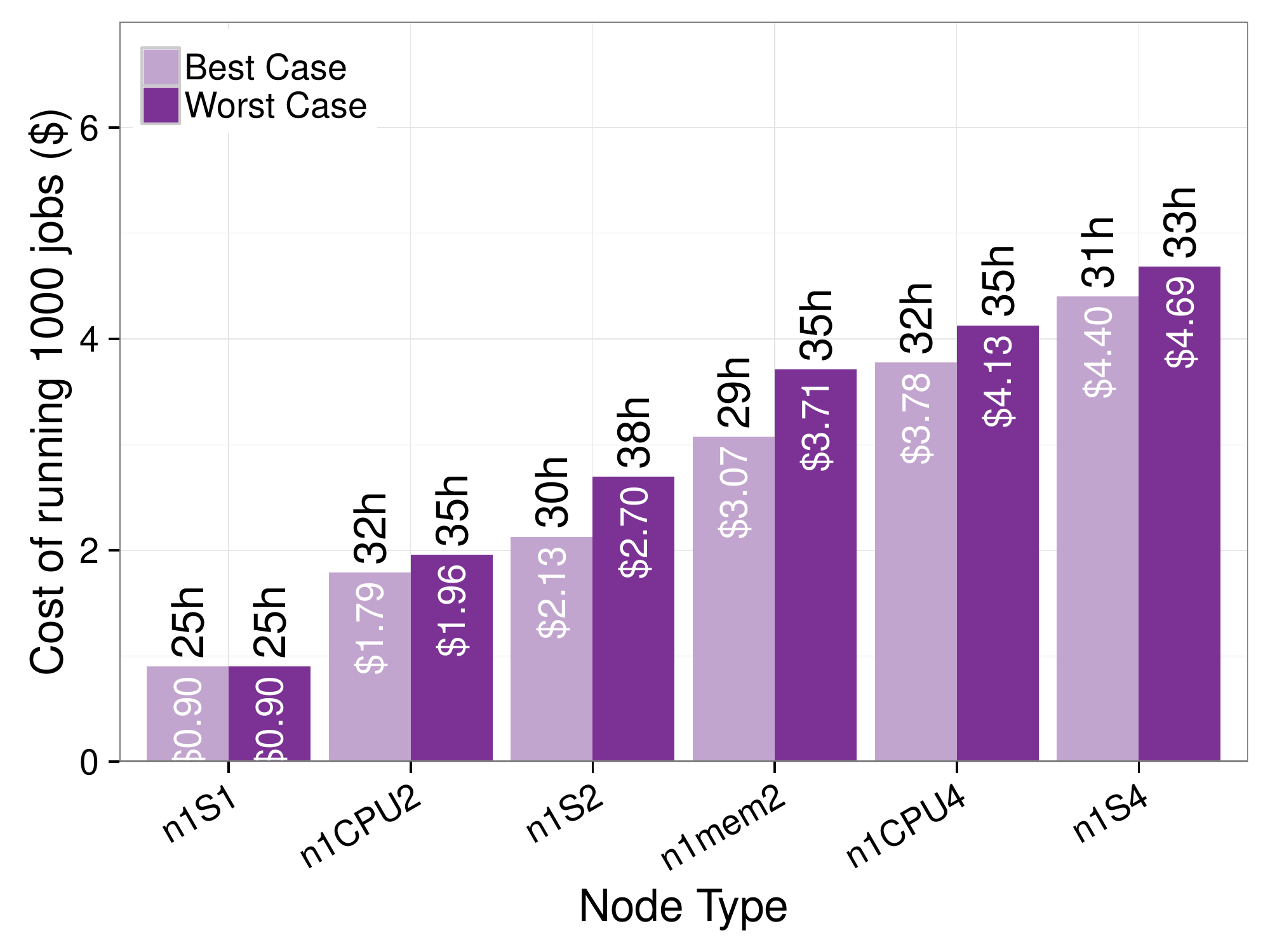}
        \caption{GCE}
	\end{subfigure}\par\medskip
	\caption{The cost of running 1000 VARD jobs.\label{fig:vard-1k}}
\end{figure*}

\begin{figure*}[tbh]
    \centering
    \begin{subfigure}{\figwidth}
        \includegraphics[width=\figwidth, trim=0cm 0.2cm 0cm 0cm, clip]{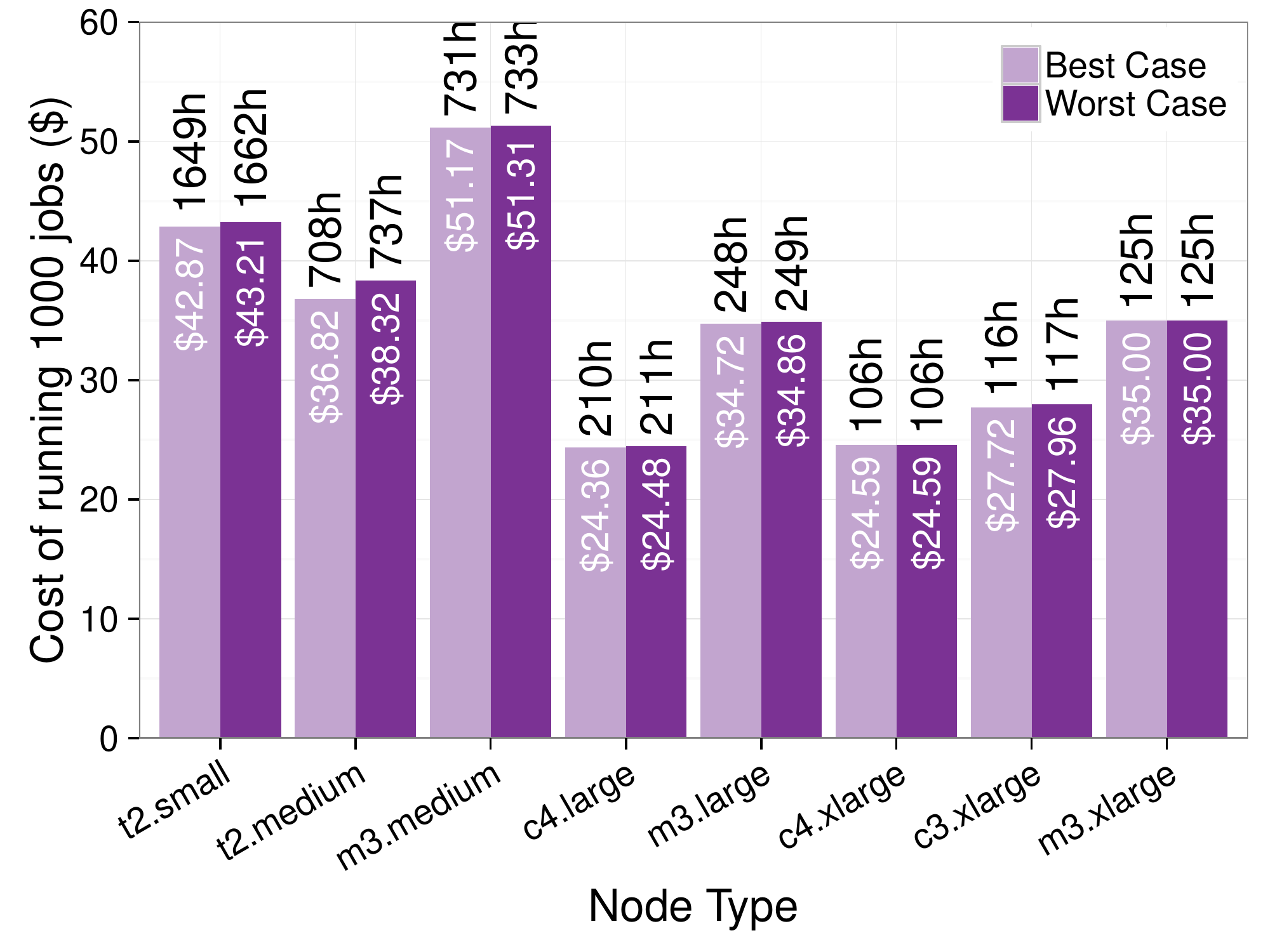}
        \caption{EC2}
	\end{subfigure}\hfill
    \begin{subfigure}{\figwidth}
        \includegraphics[width=\figwidth, trim=0cm 0.2cm 0cm 0cm, clip]{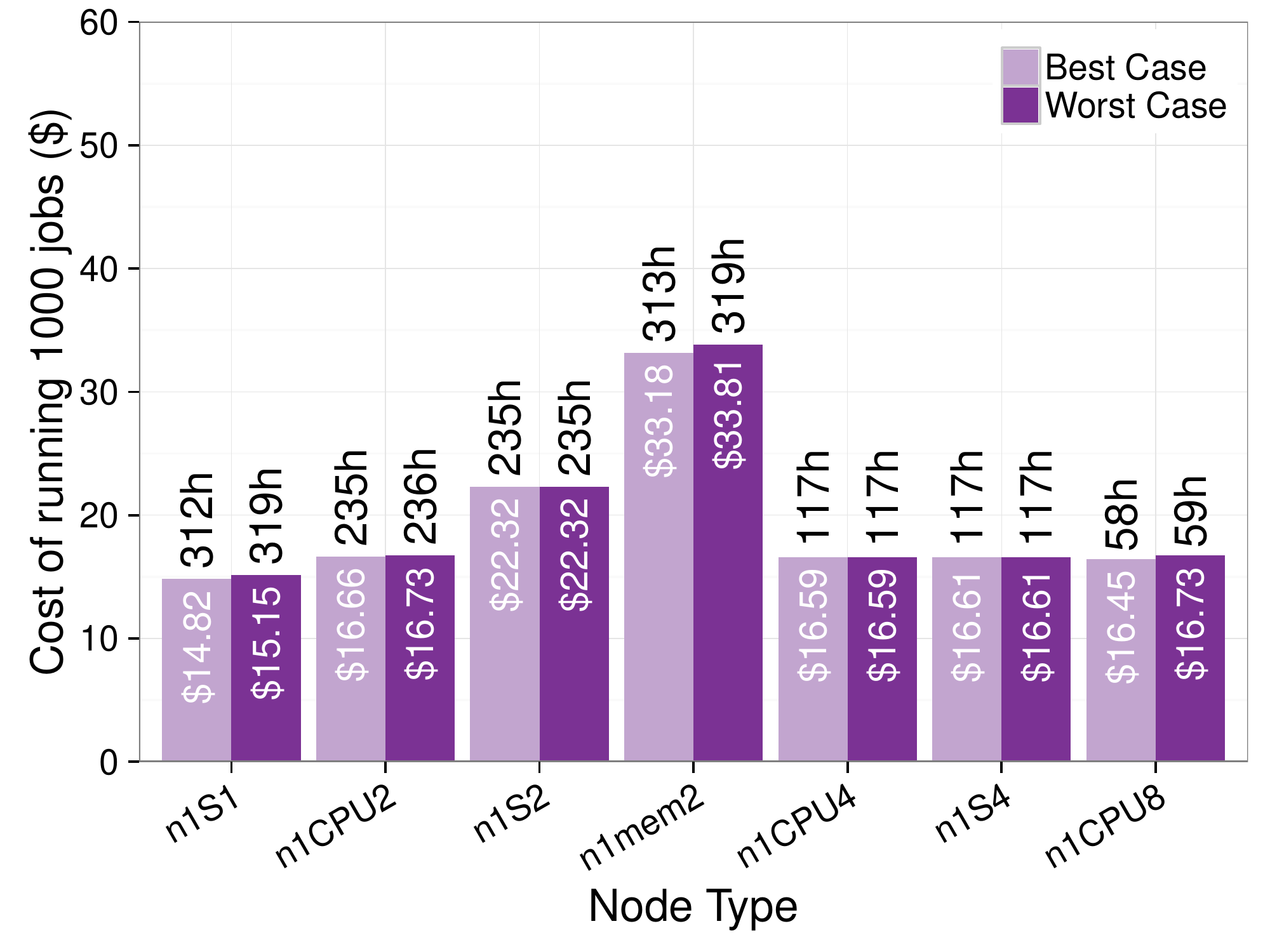}
        \caption{GCE}
	\end{subfigure}\par\medskip
	\caption{The cost of running 1000 smallpt jobs.\label{fig:smallpt-1k}}
\end{figure*}

We now try to unravel cost-effectiveness and variation therein from the perspective of IaaS users who need to execute a certain number of jobs. This is a decisive constraint for many applications where the number of sub-jobs is synonymous with work rate, aiming to process a certain amount of data points or to reduce uncertainty to a desirable level, \etc. 
For this, we again use the best and worst case scenarios defined above and apply them to the average costs of running 1,000 jobs. 
We assume that each submitted job takes the same amount of time. The results are plotted in \fig{fig:vard-1k} and \fig{fig:smallpt-1k}. We also indicate the amount of time needed for executing 1,000 jobs above each bar.

With VARD on EC2, the cheapest node \texttt{t2.small} is the most cost-effective: you spend only \$0.75 to run 1,000. This cost rises between 6 and 8 folds for the most expensive node. The amount of time reveals interesting facts as well. One can spend 27-29 hours to run 1,000 jobs on \texttt{t2.small} as opposed to 20-27 hours on \texttt{c4.xlarge} or 24-32 hours on \texttt{m3.large}. 
In effect, the user would pay much more cost for an uncertain reduction in execution time. 
\texttt{t2.medium} seems to be by far the most balanced in terms of cost and execution time: \$1.25 for 22-24 hours, \ie almost 3-5 times cheaper than the expensive nodes and with a fairly certain and acceptable execution time.

For VARD on GCE, it is both time and cost effective to use the cheapest instance type \texttt{n1S1},  which can finish 1,000 job runs for just \$0.90 in exactly 25 hours. All other instance types are more expensive in time and cost.

For smallpt, the trend of quite predictable performance is observed as seen in \S\ref{sec:results:hours} with some additional observations. For instance, EC2 \texttt{c4.xlarge} is providing the best cost:hour ratio compared to other instances. It can run 1,000 jobs for the same cost as with \texttt{c4.large}, but in nearly half the time. The same trend is noticed in \texttt{m3.xlarge} and \texttt{m3.large}, where the cost for both is very similar but the former only needs half the time to finish 1,000 jobs. 

Interestingly, the cost of running 1,000 smallpt jobs on GCE are almost equal on all instances except \texttt{n1S2} and \texttt{n1mem2}. In terms of time, \texttt{n1CPU8} (the most expensive per hour) takes only 58-59 hours which is less than half of the time needed on the next fastest instance type (\texttt{n1CPU4} and \texttt{n1S4}). Thus, \texttt{n1CPU8} is by far the best instance to use in this case. This is a clear example illustrating that it is not always the case that the cheapest node would be the most cost- or time-effective. 

smallpt also helps us draw a start contrast between the two studied CSPs. Within the instance types we studied, GCE seems to outshine EC2 for executing 1,000 smallpt job runs. Comparing \texttt{n1S2}, which the second least cost- and time-effective GCE instance, to its EC2 counterparts: it is of equivalent performance and cost to \texttt{c4.large} but much cheaper than \texttt{m3.large}. Furthermore, the general purpose GCE instance types extremely outperform the general purpose EC2 counterparts.



\section{Discussion}
\label{sec:disc}


We now reflect on the presented results and distill a number of learned lessons regarding the current state of IaaS instances.


\subsubsection*{Lesson 1: \textit{An IaaS instance does not always do what it says on the tin}}

An overarching outcome from our results is that selecting an instance type based solely on what its virtual hardware specifications are is an error-prone decision making strategy. Instances might offer different specifications but perform very similarly (\eg GCE's \texttt{n1S4} and \texttt{n1CPU4}) and vice versa (\eg EC2 \texttt{m3.large} and GCE \texttt{n1S2}). Performance variation within each instance and different pricing levels further exacerbates the decision process. 

This is a very significant outcome as it undermines a large host of previous works that allocate cloud resources based only on clock speeds and GBs of RAM. Our results demonstrate quite clearly that such model-based scheduling is, at best, na\"ive and sub-optimal. Attempts to draw up a rule set to formalise instance selection would fail to account for inexplicable deviances such as those observed with EC2 \texttt{m3.medium}. This is perhaps something that only the CSPs themselves can explicate with any confidence. Instead, deeper understanding of how different instances really perform is needed. There is already some work based on live benchmarking (\eg \cite{Verghese2013Cloudbench}) and subsequent machine learning-based allocation (\eg \cite{Matrix,Samreen2016Daleel}), and we call for future research to develop further in this direction.

\subsubsection*{Lesson 2: \textit{Superficial application profiling is insufficient}}

The other side of selecting instances based on their specifications is restricting an application to a certain type / class of cloud instances based on its stereotypical profile. We have observed how a memory-intensive application performs very poorly on a number of high-memory instances, and instead performs rather well on cpu-optimised ones.

This further confounds users wanting to optimise their cloud workloads as it essentially expands the selection space. Accordingly, this gives more impetus to use automated methods (such as those using machine learning) for the exploration of a wider search space of CSPs and their instance types. Such methods need to be adaptive in order to be able to tailor the selection process for each particular workload and its sensitivities such as those triggered by change in input parameters and/or data. 
Most big data analytic applications are composed of complex algorithms and are not only data intensive but also compute intensive. 
These applications are based on numerous data processing techniques and vary in execution behaviour and resource requirements and so the performance of different instances cannot be unified across these applications. 
Many data processing applications run for a long time and have recurring behaviour and selection of unsuitable configuration can incur excessive cost and time.

\subsubsection*{Lesson 3: \textit{Decision making is heavily driven by user constraints}}

This seemingly obvious principle is surprisingly absent from many related works that reduce cloud deployment to a simple optimisation scheduling problem. We observed how the decision to run one application (say VARD) on a single IaaS (say EC2) would change completely based on what the user's constraints. In this particular case, they could maximise the chance of running as many jobs as possible during a certain period of time by choosing \texttt{t2.medium}. If they wanted to run a certain batch of jobs and were on a budget, they would choose \texttt{t2.small}. If, instead, they wanted this to be done as soon as possible and the budget was not the restricting factor, they would be better off choosing \texttt{c4.xlarge}. These are only a basic set of constraints; there would be other sets of functional and non-functional constraints that would heavily influence the decision making process.

As such, any automated process needs to allow users to express such constraints and different combinations thereof, and be able to take said constraints in consideration when forming an instance selection method.

\subsubsection*{Lesson 4: \textit{Comparing across providers is complicated}}

This point relates to all the previous ones. There are significant differences between the instances of different CSPs, even if they appear similar on the surface. It is very difficult for users to compare across providers, especially with added uncertainty of variable instance performance. 

Both EC2 and GCE use their own flavours of well known hypervisors: Xen in the case of EC2, and KVM for GCE. From the end user perspective, these hypervisors are black boxes. The details of parallel workload on virtual machines, resource allocation algorithms, and how virtual cores are pinned to physical cores are some of the key details that are not (and probably will never be) provided by these and other IaaS CSPs to users. Consequently, IaaS users cannot perceive any collocation or interference effect on their running application. This further laments the need for automated adaptive selection processes.

A related side note: we observed EC2 performance to be more variable than GCE's. This fluctuation might be attributed to EC2's underlying hypervisor technology, Xen, which others have observed variability with \cite{6008687,Binu2011,6702584}.

\subsubsection*{Lesson 5: \textit{All is not lost}}
It seems that Pascal's quote holds true here: \textit{``It is not certain that everything is uncertain''}. Despite all the variability observed and discussed, there still remains a fair degree of certainty that is only clear once we detach ourselves from the ``book value'' of the instances in question. This, of course, comes at a cost and requires automation to achieve and also to detect. As such, there is an opportunity here to build adaptive and customisable brokers to provide such knowledge \cite{Elkhatib2016crosscloudmap}.

\section{Conclusion}
\label{sec:conc}
We carried out extensive experiments on the 2 market-leading IaaS providers, EC2 and GCE, in order to identify variances in instance performance and cost-effectiveness. We did this using 2 applications of contrasting types over a period of 7 days per provider. 
Our results indicate that instance selection incurs a considerable degree of uncertainty. Instances do not necessarily perform as well as they should based on their computational specifications. In addition, matching general application profile with instance types is suboptimal. This is especially true for running memory-intensive applications, and is more discernible in EC2 as opposed to GCE. 

Nonetheless, we still found a fair degree of confidence in instance performance albeit over large execution samples. Coupled with the sheer number of instances and their varying configurations and pricing, the search space for an optimal instance for a given application becomes substantial. However, predictability of optimal cloud instance selection can only be achieved through automated and adaptive search of such space.


\ifCLASSOPTIONcaptionsoff
  \newpage
\fi

\balance{
	\bibliographystyle{acm}
	\bibliography{bib}

\begin{thebibliography}{10}

\bibitem{Aazam2017}
{\sc Aazam, M., and Huh, E.-N.}
\newblock Cloud broker service-oriented resource management model.
\newblock {\em Transactions on Emerging Telecommunications Technologies 28}, 2
  (2017), e2937.

\bibitem{cherrypick2017}
{\sc Alipourfard, O., Liu, H.~H., Chen, J., Venkataraman, S., Yu, M., and
  Zhang, M.}
\newblock Cherrypick: Adaptively unearthing the best cloud configurations for
  big data analytics.
\newblock In {\em 14th {USENIX} Symposium on Networked Systems Design and
  Implementation ({NSDI} 17)\/} (Boston, MA, 2017), {USENIX} Association,
  pp.~469--482.

\bibitem{baron2008vard2}
{\sc Baron, A., and Rayson, P.}
\newblock {VARD2}: A tool for dealing with spelling variation in historical
  corpora.
\newblock In {\em Postgraduate conference in corpus linguistics\/} (May 2008).

\bibitem{Binu2011}
{\sc Binu, A., and Kumar, G.~S.}
\newblock {\em Virtualization Techniques: A Methodical Review of {Xen} and
  {KVM}}.
\newblock 2011, pp.~399--410.

\bibitem{Matrix}
{\sc Chiang, R.~C., Hwang, J., Huang, H.~H., and Wood, T.}
\newblock Matrix: Achieving predictable virtual machine performance in the
  clouds.
\newblock In {\em 11th International Conference on Autonomic Computing (ICAC
  14)\/} (Philadelphia, PA, June 2014), USENIX Association, pp.~45--56.

\bibitem{Elkhatib2016crosscloudmap}
{\sc Elkhatib, Y.}
\newblock {Mapping Cross-Cloud Systems: Challenges and Opportunities}.
\newblock In {\em Proceedings of the 8th USENIX Conference on Hot Topics in
  Cloud Computing\/} (June 2016), USENIX Association, pp.~77--83.

\bibitem{7027861}
{\sc Hwang, K., Bai, X., Shi, Y., Li, M., Chen, W.~G., and Wu, Y.}
\newblock Cloud performance modeling with benchmark evaluation of elastic
  scaling strategies.
\newblock {\em IEEE Transactions on Parallel and Distributed Systems 27}, 1
  (Jan. 2016), 130--143.

\bibitem{5708447}
{\sc Jackson, K.~R., Ramakrishnan, L., Muriki, K., Canon, S., Cholia, S.,
  Shalf, J., Wasserman, H.~J., and Wright, N.~J.}
\newblock Performance analysis of high performance computing applications on
  the amazon web services cloud.
\newblock In {\em Proceedings of the 2nd International Conference on Cloud
  Computing Technology and Science (CloudCom)\/} (Nov. 2010), pp.~159--168.

\bibitem{Javed201652}
{\sc Javed, B., Bloodsworth, P., Rasool, R.~U., Munir, K., and Rana, O.}
\newblock Cloud market maker: An automated dynamic pricing marketplace for
  cloud users.
\newblock {\em Future Generation Computer Systems 54\/} (2016), 52 -- 67.

\bibitem{leitner2016patterns}
{\sc Leitner, P., and Cito, J.}
\newblock Patterns in the chaos --- a study of performance variation and
  predictability in public {IaaS} clouds.
\newblock {\em ACM Transactions on Internet Technology 16}, 3 (Apr. 2016),
  15:1--15:23.

\bibitem{Cloudcmp}
{\sc Li, A., Yang, X., Kandula, S., and Zhang, M.}
\newblock Cloudcmp: Comparing public cloud providers.
\newblock In {\em Proceedings of the 10th ACM SIGCOMM Conference on Internet
  Measurement (IMC)\/} (New York, NY, USA, 2010), ACM, pp.~1--14.

\bibitem{Li2013}
{\sc Li, Z., O'Brien, L., Ranjan, R., and Zhang, M.}
\newblock Early observations on performance of google compute engine for
  scientific computing.
\newblock In {\em 5th International Conference on Cloud Computing Technology
  and Science\/} (Washington, DC, USA, 2013), IEEE Computer Society.

\bibitem{6655674}
{\sc Li, Z., OBrien, L., and Zhang, H.}
\newblock {CEEM}: A practical methodology for cloud services evaluation.
\newblock In {\em 9th World Congress on Services\/} (June 2013), pp.~44--51.

\bibitem{lucas2013multi}
{\sc Lucas-Simarro, J.~L., Aniceto, I.~n.~S., Moreno-Vozmediano, R., Montero,
  R.~S., and Llorente, I.~M.}
\newblock {\em A Cloud Broker Architecture for Multicloud Environments}.
\newblock John Wiley \& Sons, Inc., 2013, pp.~359--376.

\bibitem{6753834}
{\sc OLoughlin, J., and Gillam, L.}
\newblock Towards performance prediction for public infrastructure clouds: An
  {EC2} case study.
\newblock In {\em 5th International Conference on Cloud Computing Technology
  and Science\/} (Dec. 2013), vol.~1, pp.~475--480.

\bibitem{Ostermann2010}
{\sc Ostermann, S., Iosup, A., Yigitbasi, N., Prodan, R., Fahringer, T., and
  Epema, D.}
\newblock {\em A Performance Analysis of {EC2} Cloud Computing Services for
  Scientific Computing}.
\newblock Oct. 2009, pp.~115--131.

\bibitem{6133223}
{\sc Phillips, S.~C., Engen, V., and Papay, J.}
\newblock Snow white clouds and the seven dwarfs.
\newblock In {\em 3rd International Conference on Cloud Computing Technology
  and Science\/} (Nov. 2011), pp.~738--745.

\bibitem{Quarati2016403}
{\sc Quarati, A., Clematis, A., and D’Agostino, D.}
\newblock Delivering cloud services with {QoS} requirements: Business
  opportunities, architectural solutions and energy-saving aspects.
\newblock {\em Future Generation Computer Systems 55\/} (2016), 403 -- 427.

\bibitem{Samreen2016Daleel}
{\sc Samreen, F., Elkhatib, Y., Rowe, M., and Blair, G.~S.}
\newblock Daleel: Simplifying cloud instance selection using machine learning.
\newblock In {\em Proceedings of the Network Operations and Management
  Symposium\/} (Apr. 2016), IEEE, pp.~557--563.

\bibitem{Schad2010}
{\sc Schad, J., Dittrich, J., and Quian{\'e}-Ruiz, J.-A.}
\newblock Runtime measurements in the cloud: Observing, analyzing, and reducing
  variance.
\newblock {\em Proc. VLDB Endow. 3}, 1-2 (Sept. 2010), 460--471.

\bibitem{7037674}
{\sc Scheuner, J., Leitner, P., Cito, J., and Gall, H.}
\newblock Cloud work bench -- infrastructure-as-code based cloud benchmarking.
\newblock In {\em 6th International Conference on Cloud Computing Technology
  and Science\/} (Dec. 2014), pp.~246--253.

\bibitem{Tagg2009}
{\sc Tagg, C.}
\newblock {\em A corpus linguistics study of SMS text messaging}.
\newblock PhD thesis, University of Birmingham, July 2009.

\bibitem{Verghese2013Cloudbench}
{\sc Varghese, B., Akgun, O., Miguel, I., Thai, L., and Barker, A.}
\newblock Cloud benchmarking for performance.
\newblock In {\em 6th International Conference on Cloud Computing Technology
  and Science\/} (Dec 2014), pp.~535--540.

\bibitem{6702584}
{\sc Varrette, S., Guzek, M., Plugaru, V., Besseron, X., and Bouvry, P.}
\newblock {HPC} performance and energy-efficiency of {Xen}, {KVM} and {VMware}
  hypervisors.
\newblock In {\em 25th International Symposium on Computer Architecture and
  High Performance Computing\/} (Oct. 2013), pp.~89--96.

\bibitem{Venkataraman2016Ernest}
{\sc Venkataraman, S., Yang, Z., Franklin, M., Recht, B., and Stoica, I.}
\newblock Ernest: Efficient performance prediction for large-scale advanced
  analytics.
\newblock In {\em Proceedings of the 13th Usenix Conference on Networked
  Systems Design and Implementation\/} (Berkeley, CA, USA, 2016), NSDI'16,
  USENIX Association, pp.~363--378.

\bibitem{6008687}
{\sc Younge, A.~J., Henschel, R., Brown, J.~T., von Laszewski, G., Qiu, J., and
  Fox, G.~C.}
\newblock Analysis of virtualization technologies for high performance
  computing environments.
\newblock In {\em Proceedings of the 4th International Conference on Cloud
  Computing\/} (July 2011), pp.~9--16.

\end{thebibliography}
}

\end{document}